\documentclass[aps,prd,10pt,twocolumn,superscriptaddress,floatfix,nofootinbib,amsmath,amssymb,altaffilletter,preprintnumbers]{revtex4-1}
\usepackage[colorlinks=true,pdfstartview=FitV,breaklinks=true]{hyperref}
\usepackage[utf8]{inputenc}
\usepackage{graphicx}
\usepackage{booktabs}
\usepackage{tabularx}
\usepackage{textcomp}
\usepackage{xspace}
\usepackage{amsmath}
\usepackage{amsfonts}
\usepackage{amssymb}
\usepackage{upgreek}
\usepackage[dvipsnames,table]{xcolor}
\usepackage[normalem]{ulem}
\hypersetup{urlcolor=BlueViolet,citecolor=PineGreen,linkcolor=BrickRed}
\usepackage[official]{eurosym}
\definecolor{lightgray}{rgb}{0.9,0.9,0.9}	    
\definecolor{green}{rgb}{0,0.5,0}
\definecolor{red}{rgb}{1,0,0}
\definecolor{blue}{rgb}{0,0,0.5}

\begin{document}

\preprint{DESY-22-185}

\title{Limits on Dark Photons, Scalars, and Axion-Electromagnetodynamics with The ORGAN Experiment}

\author{Ben T.~McAllister}
\email{ben.mcallister@uwa.edu.au}
\affiliation{QDM Laboratory, Department of Physics, University of Western Australia, 35 Stirling Highway, Crawley WA 6009, Australia.}
\affiliation{Centre for Astrophysics and Supercomputing,
Swinburne University of Technology, John St, Hawthorn VIC 3122, Australia}
\author{Aaron Quiskamp}
\email{aaron.quiskamp@research.uwa.edu.au}
\affiliation{QDM Laboratory, Department of Physics, University of Western Australia, 35 Stirling Highway, Crawley WA 6009, Australia.}
\author{Ciaran A.~J.~O'Hare}
\affiliation{School of Physics, Physics Road, The University of Sydney, NSW 2006 Camperdown, Sydney, Australia}
\author{Paul~Altin}
\affiliation{ARC Centre of Excellence For Engineered Quantum Systems, The Australian National University, Canberra ACT 2600 Australia}
\author{Eugene N.~Ivanov}
\affiliation{QDM Laboratory, Department of Physics, University of Western Australia, 35 Stirling Highway, Crawley WA 6009, Australia.}
\author{Maxim Goryachev}
\affiliation{QDM Laboratory, Department of Physics, University of Western Australia, 35 Stirling Highway, Crawley WA 6009, Australia.}
\author{Michael E.~Tobar}
\email{michael.tobar@uwa.edu.au}
\affiliation{QDM Laboratory, Department of Physics, University of Western Australia, 35 Stirling Highway, Crawley WA 6009, Australia.}

\date{\today}

\begin{abstract}
Axions are a well-motivated dark matter candidate, with a host of experiments around the world searching for direct evidence of their existence. The ORGAN Experiment is a type of axion detector known as an axion haloscope, which takes the form of a cryogenic resonant cavity embedded in a strong magnetic field. ORGAN recently completed Phase 1a, a scan for axions around 65 \textmu eV, and placed the most stringent limits to date on the dark matter axion-photon coupling in this region, $|g_{a\gamma\gamma}|\leq 3\times 10^{-12}$. It has been shown that axion haloscopes such as ORGAN are automatically sensitive to other kinds of dark matter candidates, such as dark photons, scalar field/dilaton dark matter, and exotic axion-electromagnetic couplings motivated by quantum electromagnetodynamics. We compute the exclusion limits placed on these various dark matter candidates by ORGAN 1a, and project sensitivity for some future ORGAN phases. In particular, the dark photon limits are the most sensitive to date in some regions of the parameter space.
\end{abstract}

\maketitle

\section{Introduction}
The nature of dark matter is one of the most elusive mysteries in physics. It is thought to comprise the majority of the matter in our Universe \cite{Planck:2018vyg}, but its specific composition remains unknown. Although the evidence for dark matter is compelling \cite{rubin_1982,rotation_curve_1991,gravity_lensing_1998,lensing_2017, Markevitch_2004,Bertone:2016nfn}, little is known about its particle identity except that it is feebly interacting, massive and stable over timescales comparable to the age of our Universe. The search for dark matter has been underway for several decades, and includes searches for a host of potential candidates such as axions, axion-like particles and Weakly-interacting-Massive-Particles (WIMPs). 

The quantum chromodynamics (QCD) axion in particular has strong independent theoretical motivation, and interest in it as a dark matter candidate in recent times has been steadily growing~\cite{Sikivie1983,Sikivie1983b,Preskill1983,Svrcek_2006,Arvanitaki10,Higaki_2013,Baumann16,Co2020,Co2020b,Co2021,Oikonomou21,Sikivie2021,Sokolov:2021uv,DILUZIO20201,Rodd2021}. Accordingly, there is an influx of experiments attempting to directly detect QCD axion dark matter via its expected coupling to electromagnetism~\cite{Irastorza:2018dyq,Adams:2022pbo}. The coupling of axions to two photons is commonly probed using resonant cavity detectors known as axion haloscopes, first proposed by Sikivie \cite{Sikivie1984}.  In a typical haloscope, a resonant cavity is immersed in a strong DC magnetic field, which will convert galactic halo axions into detectable photons. If the generated photon frequency overlaps with the frequency of a geometrically suitable resonant cavity mode, the photon will be trapped in the resonator, and can be detected. Typical haloscopes probe the axion-photon coupling, denoted by $g_{a\gamma\gamma}$, and attempt to either discover an axion as dark matter via this coupling, or place exclusion limits on the strength of this coupling over a specific mass range. 

Whilst haloscopes have long been used for axion detection in this way, it can be shown that they are simultaneously sensitive to other types of axion-Standard Model coupling, and to other dark matter candidates entirely. Of recent interest in the field are the additional axion-electromagnetic couplings which arise from quantum electromagnetodynamics (QEMD)~\cite{Sokolov:2022fvs,TobarQEMD22}, the dark photon \cite{Nelson:2011sf,Arias:2012az,Fabbrichesi_2021,Jaeckel_2013,ohare_handbook}, and scalar-field/dilaton dark matter~\cite{SDMHaloscope}. See the recent review of Ref.~\cite{Antypas:2022asj} for a summary of models and searches.

For each of these dark matter candidates, the relationship between the unknown particle mass and the generated photon frequency in a typical haloscope which employs a DC magnetic field is given by
\begin{equation}
hf_{\rm DM} = m_{\rm DM} c^2 + \frac{1}{2}m_{\rm DM} v^2_{\rm DM}.
\end{equation}
Here $f_{\rm DM}$ is the frequency of the generated photon, $m_{\rm DM}$ is the dark matter particle mass, and $v_{\rm DM}$ is the dark matter speed with respect to the laboratory frame, expected to be approximately $10^{-3}c$ for standard dark matter halo models~\cite{Evans:2018bqy}. This relationship is simply a statement of conservation of energy in the dark matter-photon conversion process.

Since the mass of the dark matter particle is unknown,  the frequency of the generated photon is unknown, and haloscopes must tune their resonant frequency to ``scan" for the unknown rest mass.  
Additionally, the dark-matter mass is only weakly constrained by observations and theory, and may lie within a span of several orders of magnitude. This necessitates a range of experiments probing different frequency bands and equivalently, different mass ranges. Motivated primarily by the search for axions, there are several haloscope experiments around the world today, either operational, or in various stages of commissioning, these include ADMX~\cite{Bartram:2021ysp}, BREAD~\cite{BREAD:2021tpx}, CAPP~\cite{Yi:2022fmn}, DM-Radio~\cite{DMRadio:2022pkf}, GrAHal~\cite{Grenet:2021vbb}, HAYSTAC~\cite{HAYSTAC:2018rwy,HAYSTAC:2020kwv}, MADMAX~\cite{MADMAX2017,MADMAX2020}, ORGAN~\cite{Quiskamp2022,McAllister201767}, QUAX~\cite{Alesini:2019ajt}, RADES~\cite{CAST:2020rlf}, TASEH~\cite{Chang:2022hgj}, WISPLC~\cite{Zhang:2021bpa} and others. Despite being motivated by the standard axion-photon coupling, these experiments are sensitive to the other candidates outlined here, in their associated frequency or mass range.

The Oscillating Resonant Group AxioN (ORGAN) Experiment
is a microwave cavity haloscope hosted at the University of Western Australia which has recently completed a scan, setting an upper limit on axion-photon couplings $|g_{a\gamma\gamma}| \leq 3 \times 10^{-12}$ over the mass range between $63.2-67.1 \,$\textmu eV~\cite{Quiskamp2022}.  We will discuss the re-casting of this limit into limits on other dark matter-Standard Model couplings.

\section{Alternate Dark Matter Candidates}
The typical axion-photon coupling is governed by the Lagrangian term,
\begin{equation}
\mathcal{L} \supset-\frac{1}{4}g_{a\gamma\gamma}\, a(t) \, F^{\mu \nu} \widetilde{F}_{ \mu \nu}\, ,
\end{equation}
where $F^{\mu \nu}$ and $\widetilde{F}^{\mu \nu}$ are the electromagnetic field strength and its dual, and $a(t)$ is the oscillating axion field. The power extracted on resonance from a resonant cavity haloscope due to axion-photon conversion is,
\begin{equation}
P_{a \rightarrow \gamma}=\left(g_{a\gamma \gamma}^2 \frac{\rho_{a}}{m_{a}}\right)\left(\frac{\beta}{1+\beta} B_0^2 V C Q_{\mathrm{L}}\right) \, ,
\label{axion_power}
\end{equation}
(up to some dimensionful constants). The axion-photon coupling, $g_{a\gamma\gamma}$, is the primary parameter experiments aim to probe or constrain. 
The axion mass, $m_a$ sets the frequency of the field oscillations and hence the signal, while the axion density, $\rho_a$, sets the amplitude of those oscillations locally. The latter is usually assumed to be equal to the estimated dark matter density in the galaxy $\rho_a = \rho_{\rm DM} \approx 0.45$~GeV~cm$^{-3}$, and we assume the same when considering other dark matter candidates. Finally, we can see that the electromagnetic power scales with the square of the applied magnetic field $B_0$, the cavity volume $V$, the mode-dependent form factor $C$, the loaded quality factor $Q_L$, and the resonator coupling coefficient $\beta$. Parameters set by nature and by the experimentalist have been distinguished with parentheses. $C$ is given by
\begin{equation}
C = \frac{\left(\int{\vec{E}\cdot \vec{B}_0}~\mathrm{d}V\right)^2}{\left(\int{\vec{B}_0\cdot \vec{B}_0}~\mathrm{d}V\right)\left(\int{\epsilon\vec{E}\cdot \vec{E}}~\mathrm{d}V\right)}.
\label{C}
\end{equation}
Here, $\vec{E}$ and $\vec{B}$ are the resonant cavity mode fields, $\vec{B}_0$ is the applied external DC magnetic field, $\epsilon$ is the permittivity of the medium, and $V$ refers again to the volume of the resonator. 

We will now discuss the power expected to be produced in a haloscope by the other dark matter candidates in turn, and see how existing limits on the axion-photon coupling can be extended to limit the other couplings.

\subsection{Dark Photon Dark Matter}
The dark photon is a dark matter candidate which arises from a simple extension to the Standard Model, manifesting as the gauge boson associated with an added Abelian U(1) symmetry \cite{holdom_1986}. Below the electroweak scale, the dark and Standard Model photons remain coupled by a so-called kinetic mixing term. The presence of this term implies oscillations between the two with some strength given by a small dimensionless number $\chi$ which itself is often called the kinetic mixing parameter. Dark photons can also be straightforwardly given a mass, using standard techniques for making gauge fields massive like the Higgs or Stueckelberg mechanisms. 

Given that dark photons arise from an extremely minimal extension of the Standard Model, have suppressed couplings, and can be massive, they are prime candidates for dark matter. The only remaining ingredient needed is a production mechanism to explain their abundance in the universe. There are many that fit the bill. These can include even mechanisms as simple as the conventional misalignment production used also for axions~\cite{Preskill1983,Sikivie1983,Dine1983}, as long as there is also some non-minimal coupling to gravity ensuring the correct abundance~\cite{Arias:2012az,Graham:2015rva, Alonso-Alvarez:2019ixv}. Dark photons can also appear as the decay products of topological defects~\cite{Long:2019lwl}, from inflationary fluctuations~\cite{Graham:2015rva,Kolb:2020fwh,Ema:2019yrd,Ahmed:2020fhc,Nakai:2020cfw}, parametric resonance~\cite{Dror:2018pdh}, or via a tachyonic instability or axion portal~\cite{Agrawal:2018vin,Co:2018lka, Bastero-Gil:2018uel,Co:2021rhi,Gutierrez:2021gol}. 

If dark photons exist, they can be detected through the aforementioned kinetic mixing with Standard Model photons. The process is analogous to neutrino flavour oscillations, with the interaction described by the term \cite{ohare_handbook,Ghosh:2021ard}, 
\begin{equation}
\mathcal{L} \supset-\frac{1}{2} \chi F^{\mu \nu} F^\prime_{ \mu \nu} \, ,
\end{equation}
where $F^{\mu \nu}$ and $F^{\prime\mu \nu}$ are the electromagnetic and dark-photon field strengths, and $\chi$ is the kinetic mixing parameter which cosmological and astrophysical arguments constrains to be very small  $\chi \lesssim 10^{-10}$(see Ref.~\cite{CAHGit} for an up-to-date list of constraints).

It is predicted that dark photons would generate detectable photon signals inside axion haloscopes if conservation of energy conditions were met~\cite{Arias:2012az,Ghosh:2021ard,ohare_handbook}. Experiments (including axion haloscopes but also other dedicated experiments) search for dark photons in a similar way to haloscopes searching for axions. That is to say they aim to either detect dark photons as dark matter via their conversion to real photons which are deposited in microwave resonators, or constrain the strength of the kinetic mixing parameter, $\chi$ in the event of a non-detection.

Due to dark photon-photon kinetic mixing, the extracted signal power on resonance in an axion haloscope is given by  
\begin{equation}
P_{\gamma^\prime \rightarrow \gamma} = \left( \chi^2 m_{\gamma^{\prime}} \rho_{\gamma^{\prime}}\right) \left( \frac{\beta}{1+\beta} V \mathcal{G} Q_L  \right),
\label{DP_power}
\end{equation}
where $\rho_{\gamma^\prime}$ represents the local dark photon dark matter density, $\mathcal{G}$ is the mode-dependent geometry factor, which has the same form as $C$ except the dark photon polarisation vector replaces the direction of $B_0$, and $m_{\gamma^{\prime}}$ is the dark photon mass. The two geometry factors are related via $\mathcal{G} = C \, \mathrm{cos}^2(\theta)$, with $\theta$ denoting the angle between the magnetic field direction and the polarisation vector of the dark photon. Using Eqs.\eqref{axion_power} and \eqref{DP_power}, we can recast a limit on $g_{a\gamma\gamma}$ to one on $\chi$ using the following relation~\cite{Arias:2012az,ohare_handbook},
\begin{equation}
\chi=g_{a \gamma\gamma} \frac{B}{m_{\gamma\prime}|\cos \theta|} \, ,
\label{rescale}
\end{equation}
where we have used $m_a = m_{\gamma\prime}$ and $\rho_a=\rho_{\gamma\prime}$, and set the expressions \eqref{axion_power} and \eqref{DP_power} equal to one another. This is because in order to set an exclusion limit on $g_{a\gamma\gamma}$, an axion haloscope finds a minimum signal power threshold that can be excluded with some level of confidence. We can then use this same excluded signal power to exclude other sources of excess microwave power in the experiment, such as dark photon conversion. It should be noted that vetoing axion signal candidates through changing the DC magnetic field precludes axion exclusion limits from being reinterpreted as dark photon limits, since a potential dark photon signal has no dependence on $B_0$ and would therefore be discarded in an axion search which uses a DC magnetic field veto.

The only factor in the rescaling from axion to dark photon sensitivity that requires some care is the polarisation state of the dark photon dark matter locally. This enters via the term $|\cos\theta|$, where $\theta$ is the angle between the dark photon's polarisation vector, and whichever polarisation of the electromagnetic field the experiment is sensitive to. For a cavity experiment like ORGAN using TM modes, this is the angle with respect to the B-field. This means that in the case of dark photons, the cavity form factor is no longer solely something set by the experimentalist, but is a combination of experimental factors and natural ones. Moreover, it is not fully understood what the polarisation state of dark photon dark matter should be. Reference~\cite{ohare_handbook} surveyed some possibilities, focusing on the dependence on the production scenario given that the later gravitational influence of structure formation the polarisation should be minimal. Given the wide range of equivalent production scenarios for dark photons, the polarisation state could be anything from (1) completely \textit{random} in every coherence time, to (2) \textit{fixed} over time and length scales well beyond that probed by the experiment. Fortunately, presenting results that account for this uncertainty need not be so complicated, as these two cases lie at the two extremes in terms of an experiment's potential sensitivity. We refer to them as the randomised polarisation scenario, and fixed polarisation scenario respectively. 

In the randomised case a value of $\cos^2{\theta}$ is effectively drawn at random in every coherence time and hence a statistical average can be used to recast the sensitivity: $|\cos{\theta}| = \sqrt{1/3}$ if only one polarisation direction is measurable and $\sqrt{2/3}$ if there are two (as is the case for dish-based experiments for example). 

The value of $\cos{\theta}$ in the \textit{fixed} polarisation scenario requires a more involved calculation that takes into account the duration of each measurement and the position of the experiment on the Earth. For example, in cavity haloscopes conducting short ($\sim$min) measurements at each frequency and are sensitive to one polarisation direction at a time, the value of $|\cos{\theta}| \approx \sqrt{0.0019}$ is usually the value that approximately recasts 95\% CL exclusion limits. This value grows as measurement times get longer because the rotation of the Earth effectively allows the experiment to sample more possible values for the unknown polarisation axis (see Ref.~\cite{ohare_handbook} for further details). Because in this scenario one must account for the chance of having bad luck with the alignment of the dark-photon polarisation during the measurement (e.g.~if it happened to be perpendicular to the B-field), so this scenario always leads to smaller expected signals and worse limits overall.

\subsection{Scalar Field Dark Matter}

Whilst the axion is a pseudoscalar field and a well-known dark matter candidate, its scalar field counterpart, the dilaton, is an increasingly popular candidate in its own right~\cite{SDMHaloscope,SDMClock,Campbell2021,Antypas:2022asj}. Similar to the axion, the dilaton is expected to have mass, and to interact with electromagnetism. The interaction term in the Lagrangian for the dilaton-electromagnetism coupling is given by,
\begin{equation}
\mathcal{L} \supset-\frac{1}{4} g_{\phi\gamma\gamma} \, \phi(t) F^{\mu \nu} F_{ \mu \nu}\, ,
\end{equation}
where $g_{\phi\gamma\gamma}$ is the dilaton-photon coupling constant, the analogue of $g_{a\gamma\gamma}$, and $\phi(t)$ is the oscillating scalar dark matter field.

The coupling of the dilaton takes a very similar form to the axion, and consequently, the signals expected in axion haloscopes are quite similar. However, in an axion haloscope, the photon signal is such that the generated photon electric field aligns with the applied external DC magnetic field, whereas in a dilaton haloscope, the signal is such that the generated photon magnetic field aligns with the externally applied DC magnetic field, meaning that the geometric overlap integrals, or form factors are different for axions and dilatons. Specifically, under typical haloscope conditions (an applied external DC B-field in the $z$-direction of the cavity)~\cite{SDMHaloscope},
\begin{equation}
P_{\phi \rightarrow \gamma}=\left(g_{\phi\gamma \gamma}^2 \frac{\rho_{\phi}}{m_{\phi}}\right)\left(\frac{\beta}{1+\beta} B_0^2 V C_\phi Q_{\mathrm{L}}\right) \, ,
\label{dilaton_power}
\end{equation}
where $\rho_\phi$ is the density of dilaton dark matter, $m_\phi$ is the dilaton mass, and $C_\phi$ is a form factor, similar to the axion haloscope form factor $C$. The dilaton form factors i
\begin{equation}
C_\phi = \frac{\left(\int{\vec{B}\cdot \vec{B}_0}~\mathrm{d}V\right)^2}{\left(\int{\vec{B}_0\cdot \vec{B}_0}~\mathrm{d}V\right)\left(\int{\mu^{-1}\vec{B}\cdot \vec{B}}~\mathrm{d}V\right)}\, .
\label{Cphi}
\end{equation}
Here, $\vec{E}$ and $\vec{B}$ are the resonant cavity mode fields, $\vec{B}_0$ is the applied external DC magnetic field, $\mu$ is the permeability of the medium, and $V$ refers to the volume of the resonator. We can see from the above expression, and \eqref{C} that resonant modes which are suitable for typical axion haloscopes will not be immediately suitable for dilaton haloscopes, owing to the contrasting cavity mode field requirements for a high form factor. However by comparing Eqs.~(\ref{axion_power}) and (\ref{dilaton_power}), it is possible to re-cast axion exclusion limits as dilaton exclusion limits when computing both form factors for a given mode, in a manner similar to the dark photon exclusion above. Setting the excluded signal power threshold equal, $m_a$ = $m_\phi$, and $\rho_a$ = $\rho_\phi$ we arrive at the expression
\begin{equation}
g_{\phi\gamma\gamma} = g_{a\gamma\gamma}\sqrt{\frac{C}{C_\phi}} \, .
\label{phi_rescale}
\end{equation}
We can therefore simply re-scale the limits by the ratio of the form factors. 

Existing axion haloscopes optimise $C$ at the expense of $C_\phi$. In fact, in a perfectly uniform external DC magnetic field, $C_\phi$ is zero for the TM modes typically employed in axion haloscopes, so there will be no sensitivity to $g_{\phi\gamma\gamma}$. However, it can be shown that small spatial variation in real solenoid magnetic fields can lead to small but non-zero values for $C_\phi$ in the TM modes employed in haloscopes, meaning limits can be re-cast `for free' with careful computation of the convolved form factors.

\subsection{QEMD Electromagnetic Couplings}
Recent work has suggested that axions may have additional couplings to electromagnetism, beyond the well-known $g_{a\gamma\gamma}$ coupling. In particular, in a quantum electromagnetodynamic framework, it has been proposed that axions may couple via two additional terms, denoted $g_{aAB}$ and $g_{aBB}$~\cite{Sokolov:2022fvs}. 

In the context of a resonant axion haloscope, with an external magnetic field applied, no electric field applied, and a spatial gradient of the axion field $\nabla a \approx 0$, these new terms modify Maxwell's equations as follows:
\begin{align*}
\nabla\times\vec{B}_a - \dot{\vec{E}}_a &= g_{a\gamma\gamma}\dot{a}\vec{B}_0\\
\nabla\times\vec{E}_a+\dot{\vec{B}}_a &= g_{aAB}\dot{a}\vec{B}_0\\
\nabla\cdot\vec{B}_a &= 0\\
\nabla\cdot\vec{E}_a &= 0 \, .
\end{align*}
Here $\vec{E}_a$, $\vec{B}_a$ are the axion-induced electromagnetic fields, $\vec{B}_0$ is the applied external magnetic field, and $a$ is the axion field itself. In resonant haloscopes with an applied magnetic field, the $g_{aAB}$ term creates a magnetic current analogue to the well-known $g_{a\gamma\gamma}$ term's displacement current. Implementing Poynting theorem \cite{tobar2021abraham}, it has been shown that the signal power expected on resonance in a haloscope due to this additional term is given by~\cite{TobarQEMD22},
\begin{equation}
P_{aAB}=\left(g_{aAB}^2 \frac{\rho_{a}}{m_{a}}\right)\left(\frac{\beta}{1+\beta} B_0^2 V C_{aAB} Q_{\mathrm{L}}\right),
\label{aAB_power}
\end{equation}
where $C_{aAB}$ is another form factor, coincidentally of identical form to $C_{\phi}$ as presented in Eq.(\ref{Cphi}). Consequently, limits placed by haloscopes on the dilaton coupling $g_{\phi\gamma\gamma}$ are equivalent to limits on $g_{aAB}$.

\section{Results}

\begin{figure*}[t]
    \centering
    \includegraphics[width=0.9\linewidth]{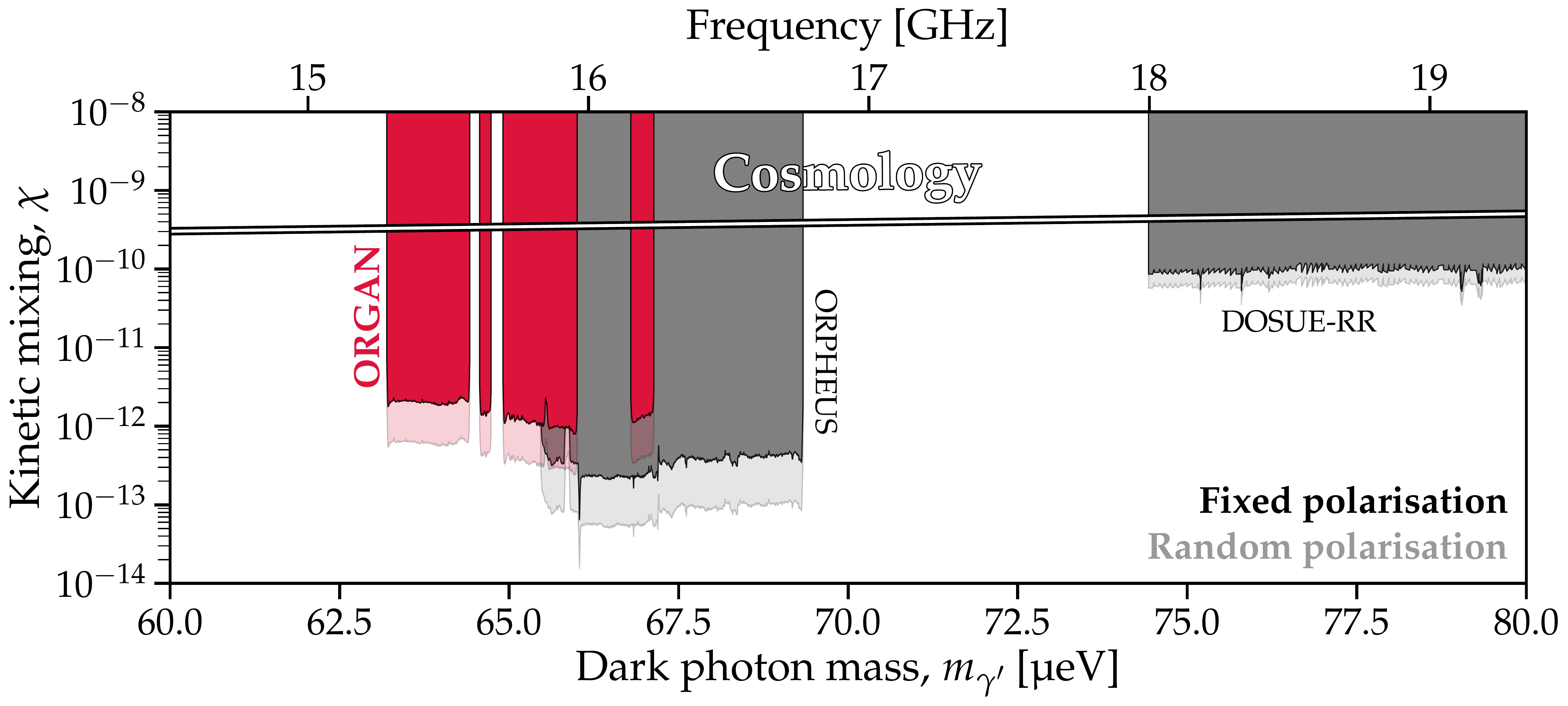}
    \caption{Close-up on the dark-photon kinetic mixing limits from ORGAN Phase 1a at 95\% CL (red), plotted against exclusion limits from ORPHEUS (90\% CL)~\cite{ORPHEUS}, DOSUE-RR (95\% CL)~\cite{DOSUE-RR:2022ise} (grey) and the cosmological bound that dark photons must meet in order to be dark matter~\cite{Arias:2012az,McDermott:2019lch,Witte:2020rvb,Caputo:2020rnx,Caputo:2020bdy} (blue). We show the limits assuming the fixed polarisation scenario as opaque, and the less-conservative random polarisation scenario as transparent (see Ref.~\cite{ohare_handbook} for the scheme to convert between the two). Limit data and up-to-date plots can be found at Ref.~\cite{CAHGit}.}
    \label{fig:DP_limits1}
\end{figure*}

\begin{figure*}[t]
    \centering
    \includegraphics[width=0.9\linewidth]{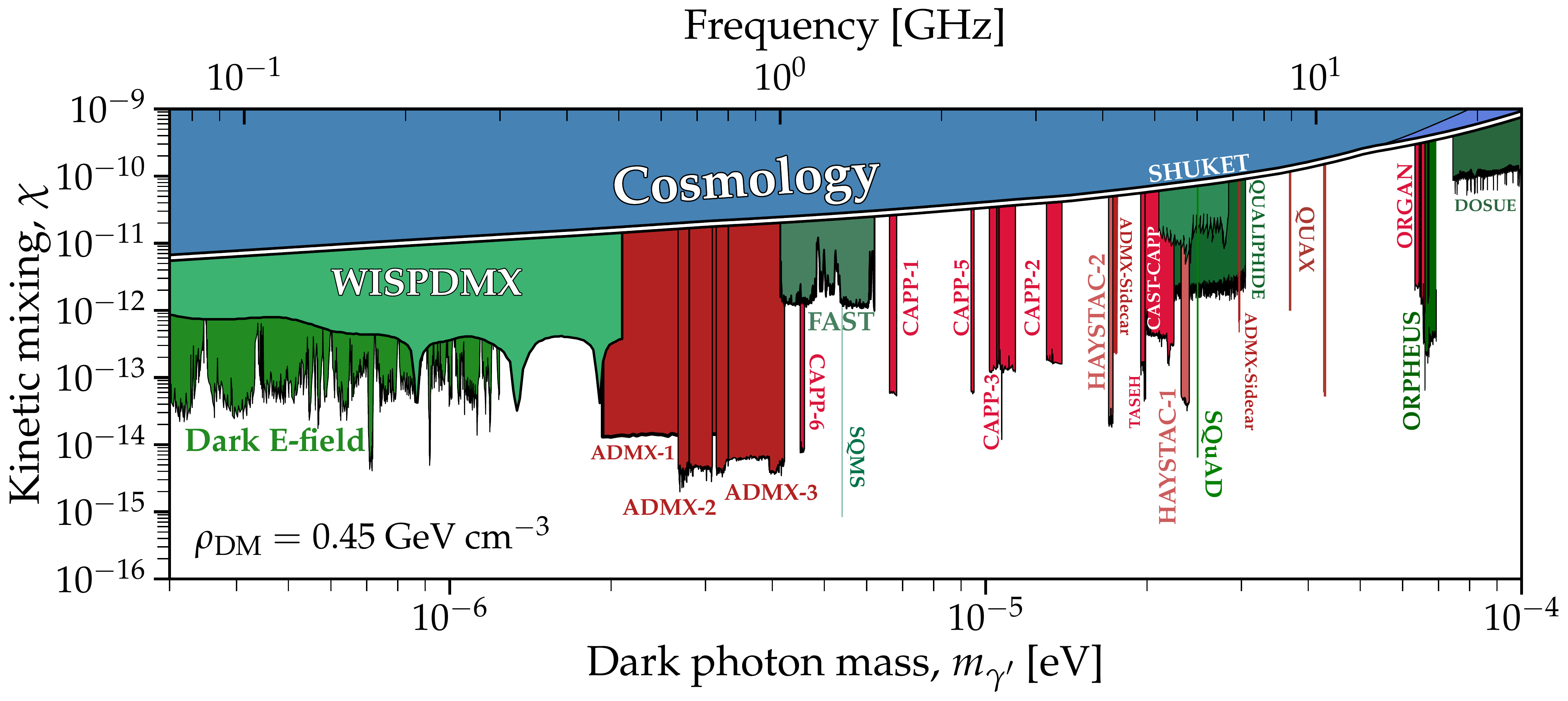}
    \caption{A wider view of the 100 MHz--10 GHz landscape of dark photon kinetic mixing limits, as they relate to the ORGAN Phase 1a limits (top right of plot). Green colours are used for limits which are set by dedicated searches for dark photons specifically~\cite{Godfrey:2021tvs,An:2022hhb,ORPHEUS,Ramanathan:2022egk,Brun:2019kak,Dixit:2020ymh,Cervantes:2022gtv,Nguyen:2019xuh}, while red colours are for re-casted axion limits~\cite{Asztalos2010,ADMX:2018gho,ADMX:2019uok,ADMX:2021nhd,ADMX:2018ogs,Bartram:2021ysp,Lee:2020cfj,Jeong:2020cwz,CAPP:2020utb,Lee:2022mnc,Kim:2022hmg,Yi:2022fmn,Adair:2022rtw,Grenet:2021vbb,HAYSTAC:2018rwy,HAYSTAC:2020kwv,Chang:2022hgj,Alesini:2019ajt,Alesini:2020vny,Alesini:2022lnp}. We assume the fixed-polarisation scenario throughout in this case, and rescale limits to the same common assumed dark matter density of 0.45 GeV cm$^{-3}$. Limit data and up-to-date plots can be found in~\cite{CAHGit}.}
    \label{fig:DP_limits2}
\end{figure*}

\begin{figure*}[t]
    \centering
    \includegraphics[width=0.87\linewidth]{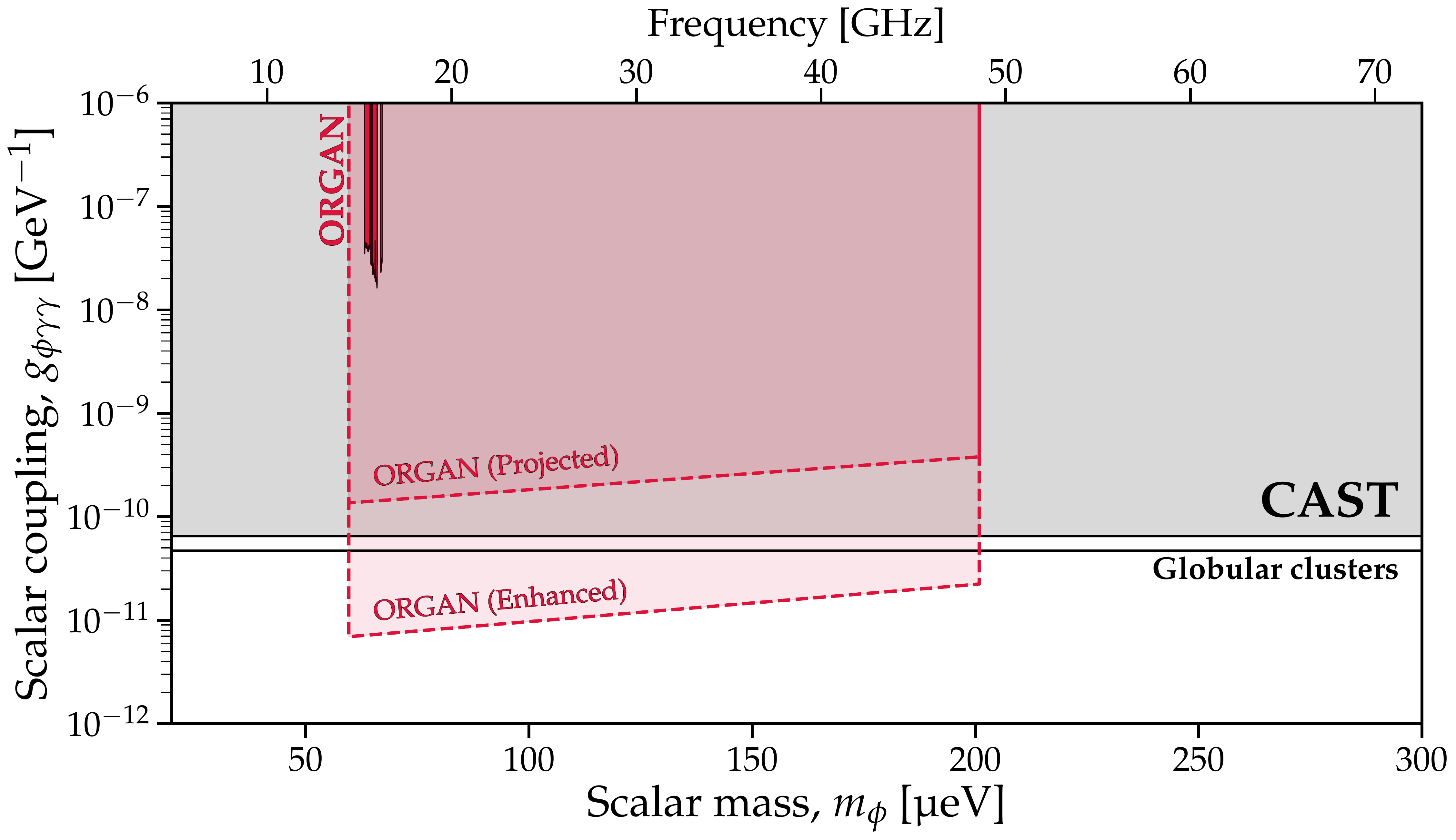}
    \caption{Limits on both the dilaton-photon coupling constant, and the axion-electromagnetic coupling parameter $g_{aAB}$ from ORGAN Phase 1a (red), presented with the limits from the CAST experiment~\cite{CAST2017,SDMHaloscope}, and the bound from considering the globular cluster $R_2$ parameter~\cite{Dolan:2022kul} (grey). We also show two projections for future ORGAN Phases with dashed lines. The higher of the two comes from assuming no enhancement to the existing experimental plans. The lower limit assumes an improvement in the relevant form factor by a factor of $\sim 300$ with one of several methods discussed in the text.}
    \label{fig:scalar}
\end{figure*}
\subsection{Dark Photon Limits}
Using Eq.\eqref{rescale}, it is straightforward to rescale the axion exclusion limits from ORGAN Phase 1a~\cite{Quiskamp2022} to limits on dark photons. ORGAN is a zenith-pointing experiment, with a mean sampling time of $\sim$1 hour at a given frequency, at a latitude of $-32^\circ$. The magnetic field amplitude was 11.5 T. Under the fixed polarisation scenario we use the code provided in~\cite{ohare_handbook}, to compute $|\cos \theta| \approx \sqrt{0.03}$ for rescaling the 95\% CL Phase 1a limits. We can adopt the value $|\cos{\theta}| = \sqrt{1/3}$ for the random polarisation scenario. For consistency with the axion community, we have taken the local dark photon dark matter density to be $0.45 \, \mathrm{GeV \, cm^{-3}}$, and have rescaled those limits that do not assume this value accordingly.

A closeup on limits from ORGAN, ORPHEUS and DOSUE-RR is shown in Fig.~\ref{fig:DP_limits1}, wherein we show the limits for both the worst-case and best-case polarisation scenarios (the limit for any arbitrary polarisation scenario will necessarily lie between these two extremes). Then in Fig.~\ref{fig:DP_limits2} we show the wider landscape of bounds over the $O(0.1$--$10)$~GHz regime assuming only the fixed polarisation scenario. We show both the limits from dedicated dark photon searches (green) and other rescaled axion searches (red). References are provided in the figure captions.

\subsection{Scalar Dark Matter Limits}
As stated in the previous section, to compute limits on $g_{\phi\gamma\gamma}$ we must compute $C_\phi$. For ORGAN 1a, to compute $C_\phi$ we perform finite element modelling to convolve the imperfectly uniform solenoid magnetic field with the $\mathrm{TM}_{010}$ cavity mode magnetic field, over the tuning range of the experimental run.

This analysis determined $C_\phi\approx 2\times 10^{-9}$, compared with $C\approx 4\times 10^{-1}$. Using Eq.\eqref{phi_rescale} we recover the limits on $g_{\phi\gamma\gamma}$ presented in Fig.~\ref{fig:scalar}. For consistency with the rest of the dark matter community, we have again taken the local dilaton dark matter density to be $0.45 \, \mathrm{GeV \, cm^{-3}}$. 

As we can see, these limits do not surpass those set by the CAST experiment~\cite{CAST2017,SDMHaloscope}, as the applied DC magnetic field is very uniform over the cavity region and the associated form factor is low. For future ORGAN phases, there are various techniques that can be employed to improve the limits on scalar dark matter. The simplest method is to offset the cavity slightly from the centre of the solenoid---this has a minimal impact on the axion form factor, $C$, but enhances the scalar form factor $C_\phi$ by a factor of $\sim 300$ due to the cavity being placed in a less uniform region of the magnet. 

Another method is to add a ring of permeable material around the outer radius of the cavity, perturbing the applied magnetic field and boosting $C_\phi$, again by decreasing field uniformity. Finite element modelling suggests that this method increases the form factor by a similar amount, a factor $\sim 300$, but at a higher penalty to the axion form factor $C$. Consequently, if an experiment were to employ this method, it would be advisable to perform separate runs for axions and dilatons.

Two projections for scalar dark matter exclusion in future ORGAN phases are shown as dashed lines in Fig.~\ref{fig:scalar}. We assume the standard, non-boosted form factor for the upper of the two limits, and for the other we assume the enhancement techniques are employed as described above.

\subsection{QEMD Axion Coupling Limits}
As discussed, the limits on the $g_{aAB}$ term which can be derived from ORGAN 1a are equivalent to the limits on $g_{\phi\gamma\gamma}$, assuming the same particle mass, dark matter density, and experimental parameters. As such, the limits and projections on $g_{aAB}$ are also presented in Fig.~\ref{fig:scalar}. We place no limits on the $g_{aBB}$ coupling term.

\section{Conclusion}
We have computed the sensitivity of The ORGAN Experiment, a high-mass axion haloscope, to various other dark matter candidates. Particularly, we compute limits from ORGAN Phase 1a on dark-photon dark matter, dilaton/scalar field dark matter, and additional axion-electromagnetic coupling terms predicted by QEMD theories. For the former, our limits are the strictest to date between 63.2 and 65.1 \textmu eV. Additionally, we show that the dilaton and new axion coupling limits of future ORGAN phases can be enhanced by implementing one of two strategies to increase the relevant form factor, which would result in the strictest limits to date on these couplings in the ORGAN mass range.

\section*{Acknowledgements}
This work was funded by the Australian Research Council Centre of Excellence for Engineered Quantum Systems, CE170100009 and Centre of Excellence for Dark Matter Particle Physics, CE200100008, and the Forrest Research Foundation. CAJO is supported by the Australian Research Council under grant number DE220100225.

\bibliography{bib}
\bibliographystyle{bibi}

\end{document}